\documentclass[twocolumn, aps, prl, superscriptaddress, floatfix]{revtex4-2}

\usepackage[utf8]{inputenc}
\usepackage{amsmath, amssymb, bm}
\usepackage{graphicx}
\usepackage{hyperref}
\usepackage{color}

\usepackage{subfigure,dcolumn,jabbrv}
\usepackage{graphicx}
\usepackage{dcolumn}
\usepackage{bm}
\usepackage{CJK}
\usepackage{url}
\usepackage{color}
\usepackage{booktabs}
\usepackage{natbib}
\usepackage{hyperref}
\usepackage{cleveref}

\usepackage[bb=ams, cal=cm, scr=boondox, frak=euler]{mathalpha}

\renewcommand{\v}[1]{\textbf{#1}}
\renewcommand{\rm}[1]{\textrm{#1}}

\newcommand{\dd}{\mathrm{d}}

\begin{document}

\title{Trace Anomaly of Cold Dense Matter Constrained by Collective Flow}

\author{Bao-An Li}
\email{Bao-An.Li@etamu.edu}
\affiliation{Department of Physics and Astronomy, East Texas A\&M University, Commerce, TX 75429-3011, USA}

\date{\today}

\begin{abstract}
The trace anomaly of dense matter, $\Delta \equiv 1/3 - P/\varepsilon$,
defined through the ratio $w \equiv P/\varepsilon$ of pressure $P$ to
energy density $\varepsilon$, quantifies deviations from conformal
symmetry and provides a dimensionless measure of the stiffness of the
equation of state (EOS) relevant for both neutron stars and heavy-ion collisions. While $\Delta(\varepsilon)$
has recently been inferred from neutron star observations, we report
the first Bayesian extraction of the trace anomaly from collective flow
observables in intermediate-energy heavy-ion collisions. By employing
transport-model simulations that explicitly decouple the cold matter
mean-field potential from thermal effects, we directly constrain the
EOS of cold dense matter. Remarkably, the trace anomaly inferred from
laboratory flow data agrees quantitatively, within $68\%$ credible
intervals, with independent astrophysical posterior bands. 
This nontrivial agreement demonstrates that heavy-ion collisions and
neutron star observations probe the same macroscopic properties in a mutually consistent way, establishing the dense-matter trace anomaly as a composition-insensitive macroscopic bridge observable across widely
different physical environments.
\end{abstract}

\maketitle

\textit{Introduction.-}Understanding the nature of supradense nuclear
matter and its EOS, encoded in the pressure--energy
density relation $P(\varepsilon)$, is a longstanding challenge shared by
nuclear physics and astrophysics.
In the gravitational regime, the $P(\varepsilon)$ relation determines the
internal structure of neutron stars through the
Tolman--Oppenheimer--Volkoff (TOV) equations
\cite{tolman1934effect,oppenheimer1939massive}.
Recent progress has shifted attention toward constraining the
dimensionless trace anomaly $\Delta \equiv 1/3 - P/\varepsilon$ \cite{Fuji22}. 
The latter is uniquely determined by the dimensionless EOS parameter $w \equiv P/\epsilon$. While $w$ is traditionally used in cosmology to classify the expansion dynamics of radiation, matter, and vacuum energy \cite{Carr01,Peeb03}, here it serves as a scale-independent measure of the medium's intrinsic stiffness. By adopting this universal parameter, we can more clearly quantify the approach to the conformal limit ($w \to 1/3$) predicted by perturbative QCD at high densities 
\cite{Bjorken83,Kurkela10,Gorda21PRL,Gorda23PRL,Gorda23,Ann23,Komo22,Braun2022,Brandt2023,Fuku2024}, and identify consistent macroscopic patterns across the vastly different energy scales of heavy-ion collisions and neutron stars.

A central difficulty in this endeavor is composition degeneracy:
macroscopic observables such as neutron star masses, radii, and tidal
deformabilities depend on the EOS $P(\varepsilon)$ but are largely
insensitive to its microscopic origin, whether nucleonic, hyperonic, or
quark matter \cite{Cai-Li}.
This degeneracy, however, suggests that $\Delta(\varepsilon)$ may serve
as a particularly robust and universal descriptor of dense matter.
While recent astrophysical analyses have begun to constrain this quantity,
an independent, non-gravitational probe is essential for establishing its
universality.

In this Letter, we provide such an independent test by performing a
Bayesian inference of the cold dense-matter trace anomaly from collective
flow observables \cite{pawel85,oll,art} in heavy-ion collisions.
Collective flow encodes the integrated response of strongly interacting
matter to pressure gradients over the full space--time evolution of the
collision. We show that the trace anomaly extracted from laboratory data at GSI
(FOPI \cite{FOPI} and HADES \cite{Ha2}) is quantitatively consistent with
constraints inferred from GW170817 and NICER observations.
This agreement identifies $\Delta(\varepsilon)$ as a unifying macroscopic
property linking femtometer-scale nuclear dynamics to kilometer-scale
neutron star structure. While trace-anomaly–related quantities have been extensively studied in
high-temperature quark–gluon plasma created in heavy-ion collisions at ultrarelativistic
energies \cite{Sen26}, QCD thermodynamics/EOS \cite{Coll,Hu10} and hadron physics \cite{Ji94}, as well as in cosmology regarding the cosmological vacuum energy \cite{Mott10}, EOS of dark energy \cite{Anto07}, quantum gravity \cite{Gab23} and cosmological inflation applications \cite{Gab25}, this work presents the first Bayesian extraction of the cold
dense-matter trace anomaly from collective flow in heavy-ion collisions
and its quantitative agreement with independent neutron star constraints.
\\

\textit{Intrinsic composition degeneracy of TOV equations and neutron star global observables.-}
Especially since GW170817, great progress has been made not only in narrowing down the $P(\varepsilon)$ band, but also more recently, in constraining the trace anomaly $\Delta(\varepsilon)$ \cite{Fuji22}.
With sound speed squared $c_s^2(\varepsilon )\equiv \frac{dP}{d\varepsilon}$, 
the ratio $w$ represents the energy-density--averaged value $\langle c_s^2(\varepsilon) \rangle$ up to $\varepsilon$ \cite{Saes24,Marc24}:
\begin{equation}
\langle c_s^2(\varepsilon) \rangle = \frac{1}{\varepsilon}\int_0^{\varepsilon} c_s^2(\varepsilon')\,\dd\varepsilon' = \frac{P(\varepsilon)}{\varepsilon} = w(\varepsilon).
\label{eq:PhiAverage}
\end{equation}
Perturbative QCD (pQCD) predicted that $\Delta$ approaches zero from above at asymptotically high densities, reflecting the approximate restoration of conformal symmetry of quark matter\,\cite{Bjorken83,Kurkela10,Gorda21PRL,Gorda23PRL,Gorda23,Ann23,Komo22,Braun2022,Brandt2023,Fuku2024}. Moreover, General Relativity (GR) requires $\Delta_{\rm GR}\geq -(0.041\sim 0.051)$ \cite{CLZ23-a,CLZ23-b,Cai:2026rzp}. 

\begin{figure}[thb]
\centering
\hspace{-1.2cm}
 \resizebox{0.56\textwidth}{!}{
\includegraphics[width=0.7\textwidth]{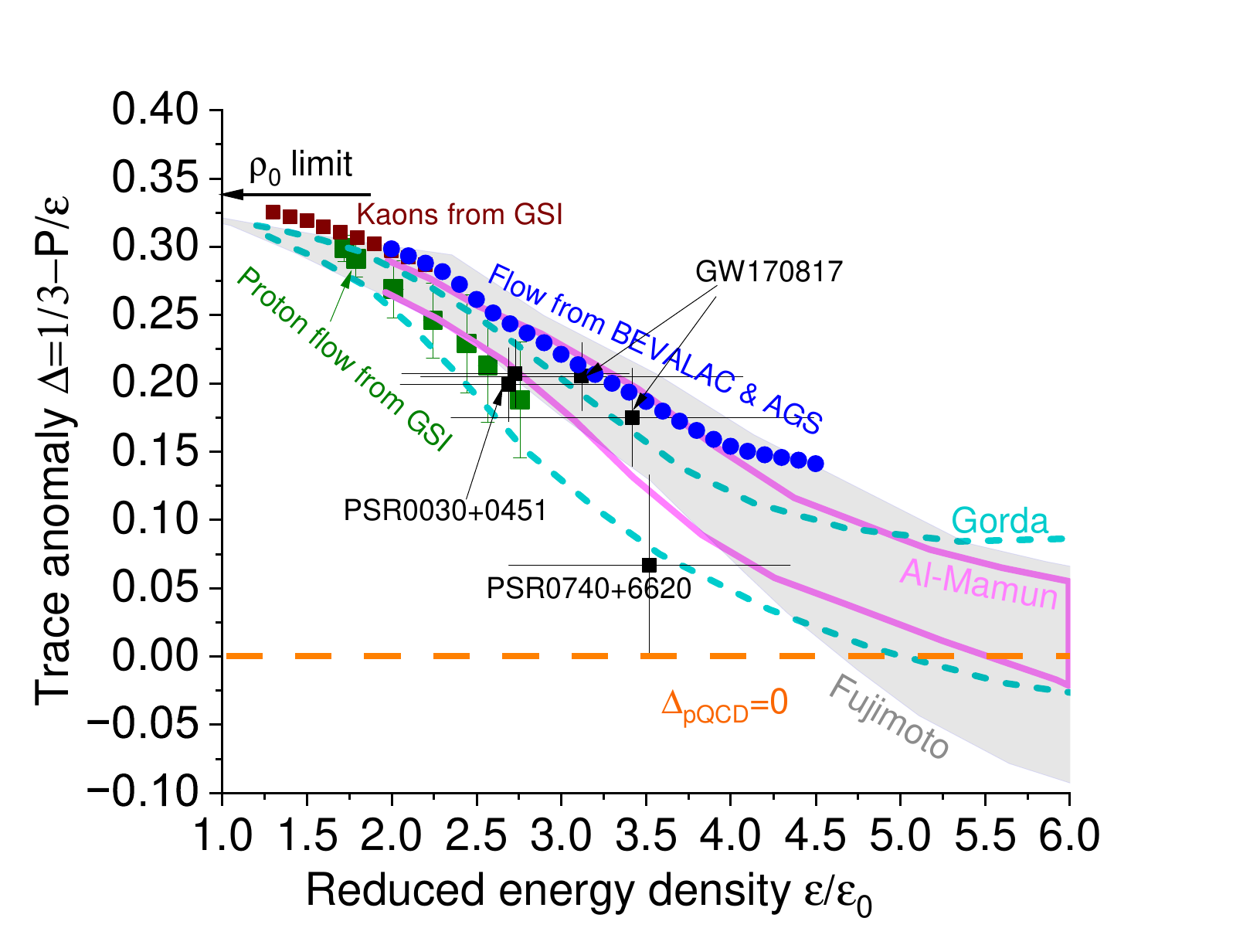}
}
\vspace{-0.7cm}
\caption{Trace anomaly $\Delta(\varepsilon)$ as a function of reduced energy density $\varepsilon/\varepsilon_0$ with
$\varepsilon_0 \simeq 150~\mathrm{MeV/fm}^3$ denotes the energy density at the saturation nucleon density $\rho_0 \simeq 0.16~\mathrm{fm}^{-3}$ of symmetric nuclear matter. Values are shown as inferred from Bayesian analyses of neutron star observational data by (1) Gorda \textit{et al.} \cite{Gorda23}, (2) Al-Mamun \textit{et al.} \cite{Mam2021}, and (3) Fujimoto \textit{et al.} \cite{Fuji22}. Also plotted are the central $\Delta(\varepsilon)$ values \cite{CL25-b} for the binary neutron stars GW170817 \cite{LIGO}, PSR~J0030+0451 \cite{Miller19,Riley19}, and PSR~J0740+6620 \cite{Salmi22}. These astrophysical constraints are compared with Bayesian-inferred $\Delta(\varepsilon)$ values (green squares) obtained from proton collective flow data measured by the FOPI and HADES Collaborations at GSI \cite{FOPI,Ha2}. For historical context, we also include $\Delta(\varepsilon)$ values obtained by converting the continuous pressure bands previously extracted through forward-modeling of kaon production \cite{Fuchs05,lynch09} and collective flow at BEVALAC and AGS energies \cite{Pawel02}. The pQCD limit and $\Delta(\rho_0)=1/3$ are indicated by the horizontal lines.} \label{DT1}
\end{figure}
Since the trace anomaly is fully determined by the EOS $P(\varepsilon)$,
one may ask why extracting it provides a more robust or less model-dependent
constraint than directly inferring the EOS itself. We address this important
question from two complementary aspects. First, the ratio $w=P/\varepsilon$ is a dimensionless quantity that
characterizes the intrinsic stiffness of the EOS. Dimensionless EOS
parameters are often more robust against microscopic modeling uncertainties
because they remove normalization scales associated with the absolute values
of pressure and energy density. A well-known example is the use of $w$ in cosmology to classify radiation, matter, and
vacuum energy independently of microscopic details
\cite{Carr01,Peeb03}. Similarly, in hot QCD thermodynamics the trace anomaly
$(\varepsilon-3P)/T^4$ is widely used to quantify interaction effects in
strongly interacting matter at temperature $T$
\cite{Coll,Hu10,Ji94}. Analogously, the ratio $P/\varepsilon$ in dense matter
provides a dimensionless, averaged measure of the EOS stiffness, as shown in
Eq.~(\ref{eq:PhiAverage}), and therefore provides a robust characterization of
the thermodynamic response of matter under extreme compression.\\

Second, the robustness of the ratio $P/\varepsilon$ can be understood analytically using the Hugenholtz--Van Hove (HVH) theorem~\cite{Hug58,Sat99}. For a zero-temperature Fermi system, $P/\varepsilon=\rho E_{\rm F}/\varepsilon-1
=E_{\rm F}/{E}-1$,
depends only on the relation between the Fermi energy $E_{\rm F}$ and the average energy per particle $E$, and is therefore insensitive to how the energy density is partitioned into kinetic and interaction contributions. This further demonstrates that the trace anomaly provides a macroscopic characterization of the EOS independent of microscopic decomposition.

It is known that the TOV equations \cite{tolman1934effect,oppenheimer1939massive} are intrinsically composition degenerate, namely, they only see the  $P(\varepsilon)$ but not its microscopic origin. Whether the pressure comes from nucleons, hyperons, quarks, two-body forces, three-body forces, short-range correlations, or other sources is irrelevant unless they uniquely change the EOS itself. While in forward modeling of neutron stars, one can use a chosen microscopic model to construct a unique EOS to predict observables (e.g., mass, radius, tidal deformation), the inverse inference from observables to microscopic models is not unique. Fundamentally, the forward modeling tells us how a given microscopic theory maps into observables. But inverse inference tells us what part of the theory survives coarse-graining. That surviving part from analyzing neutron star global observables is $P(\varepsilon)$ and its derivatives, but not the underlying Hamiltonian. Thus, neutron star observables constrain the macroscopic EOS, while each forward modeling provides one possible microscopic realization consistent with it. 

Indeed, the trace anomaly $\Delta(\varepsilon)$ inferred recently from neutron star
observations using various Bayesian frameworks—constrained by chiral
effective field theory at low densities and perturbative QCD at high
densities—yields mutually consistent results across several independent
analyses. This is illustrated by three examples in Fig.~\ref{DT1} from
Gorda {\it et al.} \cite{Gorda23}, Al-Mamun {\it et al.} \cite{Mam2021},
and Fujimoto {\it et al.} \cite{Fuji22}. These studies employ different
EOS parameterizations, priors, and statistical implementations, yet the
resulting $\Delta(\varepsilon)$ bands overlap closely within their quoted
$68\%$ credible regions. Also shown are the central values of
$\Delta(\varepsilon)$ \cite{CL25-b} for the binary neutron star merger
GW170817 \cite{LIGO}, PSR0030+0451 \cite{Miller19,Riley19} and
PSR0740+6620 \cite{Salmi22} observed by NICER, inferred without assuming any specific
microscopic EOS model from perturbative solutions of the dimensionless
TOV equations for reduced neutron star variables \cite{CLZ23-a}. Taken together, these results indicate that current neutron star observations constrain a relatively robust, coarse-grained measure of the EOS
stiffness, even though the detailed inference of $\Delta(\varepsilon)$ may depend on the adopted parameterizations and priors.
\\ 

\textit{What can be Bayesian inferred from nuclear collective flow data in
heavy-ion collisions at intermediate energies?}—
Collective flow observables \cite{pawel85,oll,art} in intermediate-energy
heavy-ion collisions are commonly interpreted within hydrodynamic or
transport frameworks in which pressure gradients drive the acceleration
of the medium
\cite{Pawel02,Sto86,Bertsch,Cas90,res97,Bass,Buss,Heinz13,Trautmann,
Sor24,Wang20}. At the level of ideal relativistic hydrodynamics this
physics \cite{oll,pawel} is encapsulated in the Euler equation governing
the collective velocity field $\vec{v}$,
\begin{equation}
(\varepsilon + P)\,\partial_t \vec{v} = - \nabla P ,
\label{eq:Euler}
\end{equation}
where $\varepsilon + P$ plays the role of an inertial mass density and
$\nabla P$ provides the driving force. Importantly, Eq.~(\ref{eq:Euler})
depends only on the macroscopic EOS $P(\varepsilon)$ and is insensitive to the microscopic mechanisms that generate the
pressure and energy density.

We emphasize that Eq.~(\ref{eq:Euler}) should be understood as a
macroscopic organizing principle rather than implying that the matter
created in intermediate-energy heavy-ion collisions behaves as an ideal
fluid. In practice, collective flow has long been analyzed using
transport approaches that incorporate finite mean free paths,
non-equilibrium dynamics, and microscopic scattering processes. Within
these models, a variety of microscopic inputs—such as the momentum dependence of the
single-particle potential, in-medium modifications of the scattering cross
sections, and different realizations of the nuclear EOS, can affect the
detailed magnitude of flow observables. Nevertheless, extensive
forward-modeling studies over the past four decades have shown that the
pressures extracted from flow data are remarkably consistent within relatively narrow bands (with respect to those extracted from neutron star properties, see, e.g., Refs. \cite{Huth21,Tsang23}) across
different transport implementations, from the pioneering work of
Ref.~\cite{Pawel02} to recent systematic surveys \cite{LeF16,Cozma24,Jerome}.
This empirical convergence indicates that collective flow is primarily
sensitive to the macroscopic relation $P(\varepsilon)$ rather than to
the specific microscopic ingredients entering the transport dynamics.

Collective flow reflects an integral response to pressure gradients over the full space–time evolution of the collision ~\cite{Li22,f2}. Rewriting Eq.~(\ref{eq:Euler}) as
\begin{equation}
\partial_t \vec{v}
\sim -\frac{c_s^2(\varepsilon)}{1 + P/\varepsilon}\,\nabla \ln \varepsilon,
\end{equation}
shows that the observed flow encodes a space--time--weighted average of $c_s^2(\varepsilon)$ and is therefore primarily sensitive to coarse-grained quantities such as $\Delta(\varepsilon)$. Since $P(\varepsilon)=\int_0^{\varepsilon} c_s^2(\varepsilon')\,d\varepsilon'$, the trace anomaly $\Delta(\varepsilon)$ encodes the EOS stiffness averaged up to $\varepsilon$. Evaluated at the maximum density reached in the collision, it therefore characterizes the effective stiffness probed by the reaction.

Within this framework, using a Gaussian-process emulator trained on an isospin-dependent
Boltzmann–Uehling–Uhlenbeck (IBUU) transport model \cite{LiBA04,BALI} for intermediate-energy heavy-ion reactions with an EOS described in End Matter and Refs. \cite{LX-HA,Li25-PRC}, we infer the trace anomaly $\Delta(\varepsilon)$. To our knowledge, this represents the first Bayesian extraction of this quantity from intermediate-energy heavy-ion
collective flow data. Future studies employing different transport
models using various EOS parameterizations, including possibly meta-model EOSs (see, e.g., Ref. \cite{Bur25}), will provide valuable cross-checks of the robustness of this inference, although such model-to-model comparisons lie beyond the
scope of the present work.

\textit{Bayesian inference of the trace anomaly and speed of sound from collective flow in heavy-ion collisions.—}
We recently performed Bayesian analyses \cite{LX-HA,Li25-PRC}
of the excitation functions of protons' directed and elliptic flow
measured by the FOPI Collaboration at beam energies from
150 to 1200~MeV/nucleon, as well as the HADES data at
1230~MeV/nucleon \cite{Ha2}, for mid-central Au+Au collisions.
The matter produced in heavy-ion collisions at intermediate beam energies 
does not behave as a perfectly ideal fluid; the transport-model framework employed here
naturally incorporates non-ideal effects such as finite mean free paths
and viscosity, reinforcing the robustness of the inferred trace anomaly
beyond the ideal-hydrodynamic limit. 

In transport simulations of heavy-ion collisions at intermediate energies, the EOS of cold nuclear matter enters through the single-nucleon mean-field potential\cite{Bertsch}. Temperature-related effects are treated separately via collision integrals. In particular, to account for uncertainties associated with particle scattering in
dense matter, we introduced an in-medium baryon–baryon scattering
cross-section (BBSCS) modification factor $X$ (relative to free-space
values) and assign it a broad prior range of $0.3 \leq X \leq 2.0$. Posterior distributions are obtained using the emulator-based Bayesian framework described in Refs.~\cite{LX-HA,Li25-PRC}. The posterior
distributions are well localized within the adopted prior ranges and are therefore primarily data-driven. Because the inference constrains the macroscopic relation $P(\varepsilon)$ rather than a specific microscopic realization, different EOS parameterizations that reproduce the same $P(\varepsilon)$ will yield essentially identical trace anomalies within the present uncertainties. Additional discussions of the role of momentum-dependent interactions and their impact on the extraction of the cold EOS are provided in the End Matter.

\begin{figure}[thb]
\centering
 \resizebox{0.55\textwidth}{!}{
\includegraphics[width=1.\textwidth]{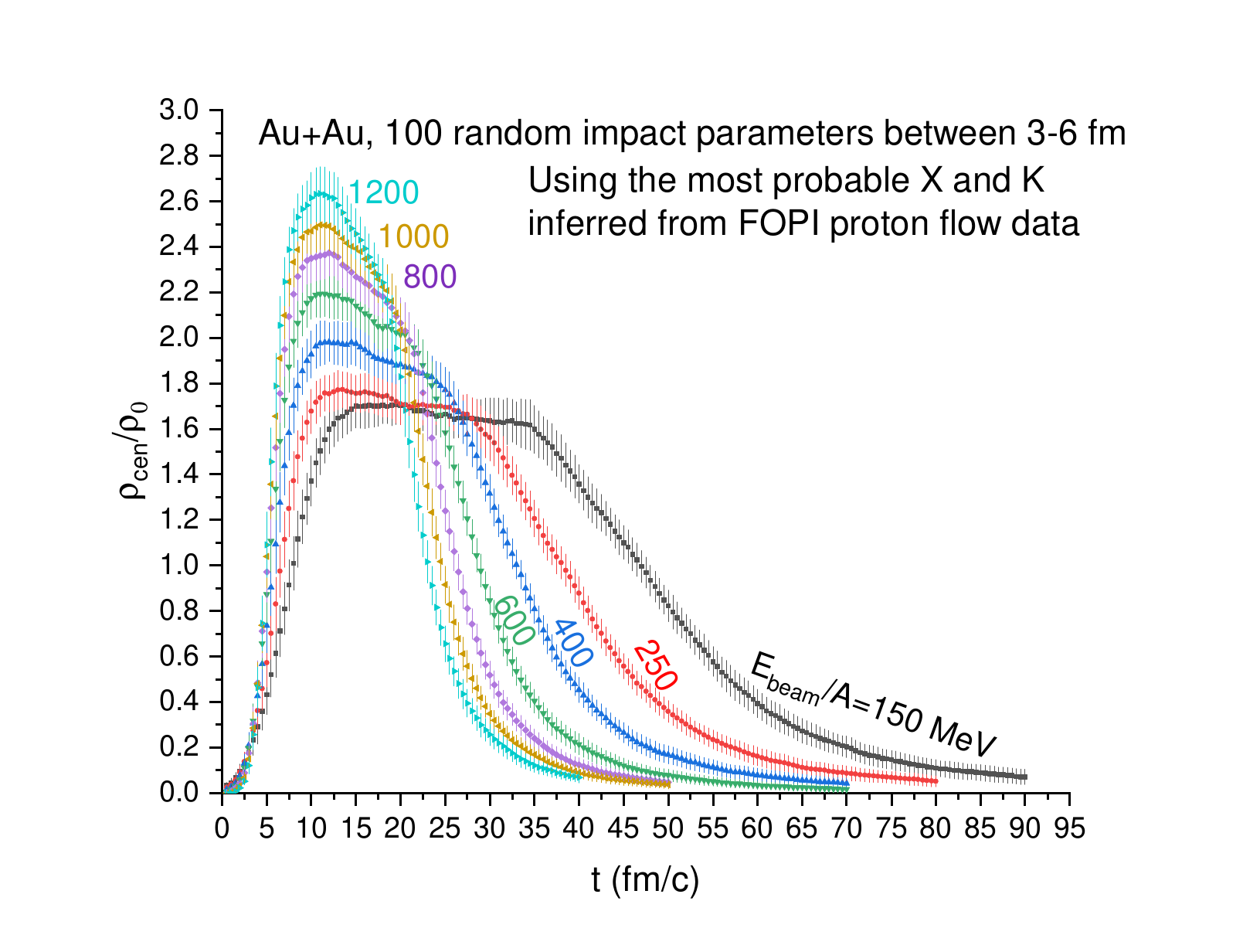}
}
\caption{Time evolution of central baryon density in mid-central Au+Au reactions with beam energies from 150 to 1200 MeV/nucleon with parameters listed in Table \ref{mean} of End Matter.
} \label{density}
\end{figure}

The time evolution of the central baryon density,
$\rho_{\rm cen}/\rho_0$, evaluated in a cubic cell of volume $1~\mathrm{fm}^3$
centered at the center of mass of the reaction system, is shown in
Fig.~\ref{density}.
For each beam energy, the maximum central density attained during the collision
is used to evaluate the trace anomaly, with the sound speed squared $c_s^2$
averaged up to this density according to Eq.~(\ref{eq:PhiAverage}).
The systematic beam-energy dependence of the maximum central density then
enables the extraction of the trace anomaly as a function of the reduced energy
density, shown as green squares in Fig.~\ref{DT1}. 

Remarkably, as shown in Fig. \ref{DT1}, the trace anomaly extracted from collective flow in
heavy-ion collisions exhibits quantitative agreement with independent
constraints inferred from neutron star observations based on
mass--radius and tidal deformability measurements. This agreement provides an independent consistency check of the EOS.
In the Bayesian analysis, the specific microscopic transport model used
in the forward simulations serves solely as a generator of training
data that spans a sufficiently broad space of EOS realizations—most
notably encompassing the full range of the incompressibility parameter
$K$—rather than as a source of model-dependent microscopic constraints.
Consequently, the inferred trace anomaly reflects EOS information that
is robust under changes of microscopic dynamics, explaining its
consistency with independently inferred neutron star constraints.

For comparison, we have also translated the $P(\rho)$ bands extracted
approximately two decades ago, from kaon production data analyzed within quantum molecular dynamics (QMD) transport
models~\cite{Fuchs05,lynch09}, and from collective flow data at
BEVALAC and AGS energies studied using both QMD- and BUU-type transport
models~\cite{Pawel02}, into the trace anomaly $\Delta(\varepsilon)$.
In making this comparison, we approximate the energy density as
$\varepsilon = M_N \rho$ where M$_N$ is the mean mass of nucleons, since the earlier constraint bands were
obtained by combining multiple analyses based on different nuclear
interactions. The resulting uncertainty introduces only a very small
horizontal shift because the combined kinetic and potential energy
contributions amount to less than $\sim 10\%$ of $M_N$ (for $K=300$ MeV)
up to $3\rho_0$. We emphasize that in our Bayesian analyses the energy
density $\varepsilon$ includes the nucleon rest-mass, kinetic, and
potential energy contributions exactly as in Eq.~(\ref{EN}) given in End Matter. These earlier constraints are broadly consistent with both the neutron star $\Delta(\varepsilon)$ bands and the results extracted here from the FOPI and HADES flow data, supporting the robustness of the trace anomaly as a measure of dense-matter stiffness.

\begin{figure}[thb]
\centering
\vspace{-0.5cm}
 \resizebox{0.55\textwidth}{!}{
\includegraphics[width=3.5cm]{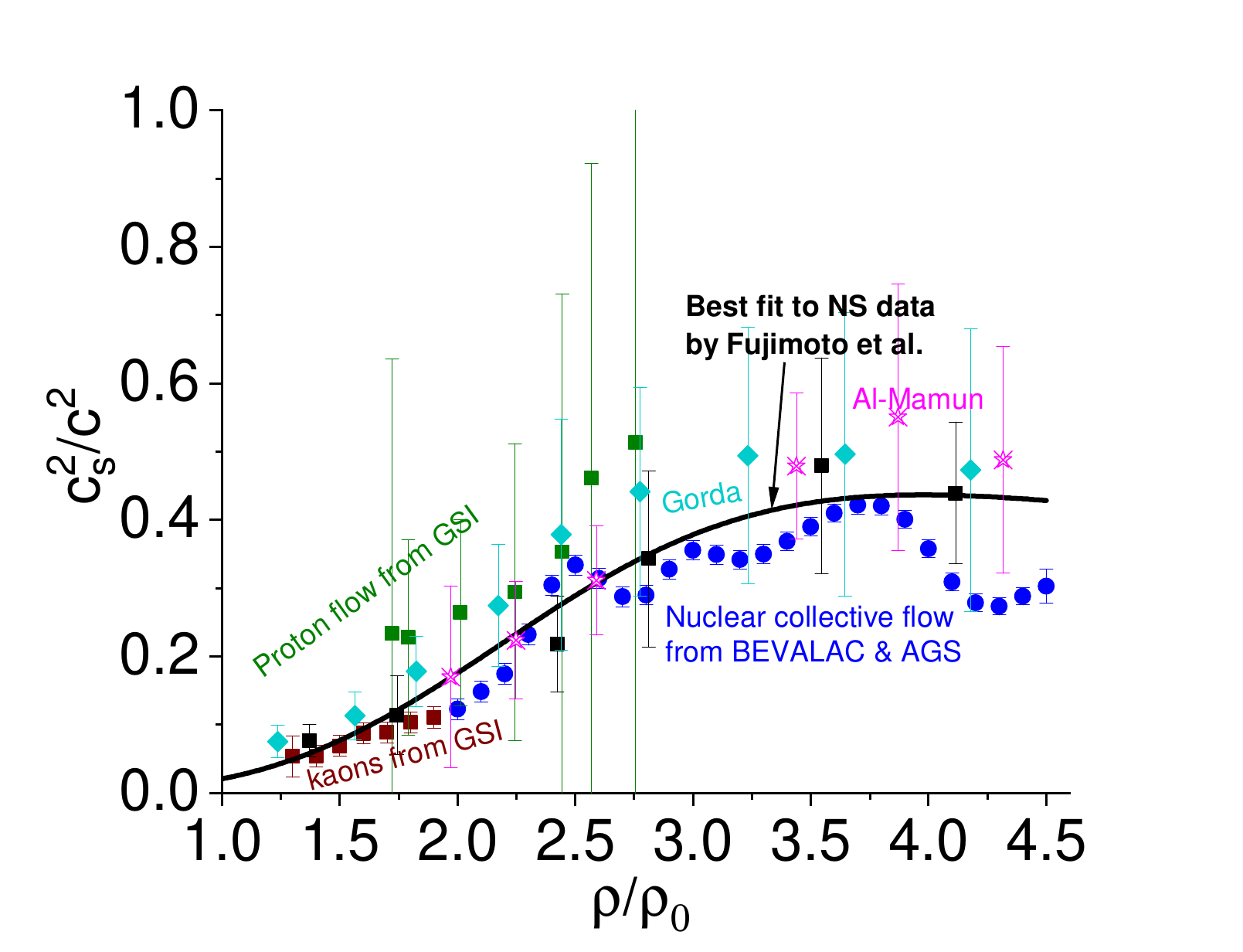}
}
\caption{Speed of sound squared derived from the trace anomaly shown in Fig. \ref{DT1} using Eq.\ref{ss_decom} in comparison with the best-fit curve inferred from neutron star observations using machine learning techniques \cite{Fuji22}.} \label{DT2}
\end{figure}

Because $P(\varepsilon)=\int_0^{\varepsilon} c_s^2(\varepsilon')\,d\varepsilon'$, one may interpret $P/\varepsilon$ as an effective average of $c_s^2$ over density based on the mean value theorem. This non-uniqueness does not affect the physical interpretation.
By studying the beam-energy dependence of the flow signal, which
corresponds to different maximum compression densities, and inverting Eq. (\ref{eq:PhiAverage}), we reconstruct the density dependence of the sound speed squared $c_s^2(\varepsilon)$.
Analytically, the density dependence of the sound speed can be written
directly in terms of the trace anomaly. Using the definition
$\Delta(\varepsilon)=1/3-P(\varepsilon)/\varepsilon$, one obtains~\cite{Fuji22}
\begin{equation}
c_s^2(\varepsilon)
\equiv \frac{dP}{d\varepsilon}
= -\bar{\varepsilon}\frac{d\Delta}{d\bar{\varepsilon}}
+ \frac{1}{3} - \Delta ,
\label{ss_decom}
\end{equation}
where $\bar{\varepsilon}\equiv\varepsilon/\varepsilon_0$.
The first term represents the derivative contribution associated with the density variation of the trace anomaly, while the remaining terms form the non-derivative part.

Figure~\ref{DT2} compares the resulting $c_s^2(\rho)$ inferred from heavy-ion collisions with constraints from neutron star observations. Despite comparatively larger uncertainties arising from the derivative
nature of Eq.~(\ref{ss_decom}) and the finite sampling of beam energies in
the GSI data, the inferred $c_s^2(\rho)$ values are consistent with the 
neutron star constraints. Overall, the mean values of $c_s^2(\rho)$ from heavy-ion collisions cluster
closely around the best-fit sound-speed profile obtained independently
from neutron star observations using machine-learning techniques~\cite{Fuji22}.

Taken together, neutron star observables and heavy-ion collective flow provide independent probes of dense matter that are primarily sensitive to the macroscopic EOS. 
Their convergence indicates a consistent macroscopic EOS over the explored density range, establishing the trace anomaly as a bridge observable linking terrestrial experiments and astrophysical observations.

\textit{Conclusions.--}
In summary, we have demonstrated that collective flow observables in
intermediate-energy heavy-ion collisions enable the first Bayesian
extraction of the cold dense-matter trace anomaly, $\Delta(\varepsilon)$,
and the corresponding EOS parameter $w \equiv P/\varepsilon$.
Although these reactions create hot and dynamically evolving systems, our transport simulations
enable a practical decoupling of the cold-matter EOS—encoded in the
nucleon mean-field potential—from thermal effects treated through
collision terms. Within this framework, collective flow data constrain
the macroscopic stiffness of the cold dense-matter EOS. Because
collective flow develops continuously during the reaction and is
generated by pressure gradients integrated over the space--time
evolution of the system, the inferred trace anomaly at a given beam
energy represents an effective average of the EOS stiffness over
densities up to the maximum compression reached in the collision.

The trace anomaly inferred from these laboratory experiments shows
remarkable quantitative consistency with independent constraints
derived from global neutron star observables. This agreement is nontrivial,
given the very different data sets, priors, and inference frameworks. It indicates that collective flow and neutron star global observables probe the same macroscopic properties of cold dense matter despite their very different dynamical conditions. Our results demonstrate that collective flow provides a robust
laboratory probe of the macroscopic EOS that is complementary to
neutron star measurements. By exploiting the beam-energy dependence of
collective flow, which corresponds to different maximum compression
densities reached in the collision, the trace anomaly can be mapped
over a range of energy densities and used to reconstruct the density
dependence of the EOS stiffness. Future high-precision experiments at
next-generation heavy-ion facilities, together with improved
constraints from advanced X-ray timing and gravitational-wave
observations of neutron stars, will allow stringent tests of the
dense-matter trace anomaly and further refine constraints on the EOS
over a broad range of densities. These developments will further establish the trace anomaly of cold dense matter—constrained by collective flow—as a bridge observable linking terrestrial heavy-ion experiments and neutron star observations.
\\

\textit{Acknowledgement.-}
We thank B.J. Cai, X. Grundler, J. Margueron, W.J. Xie, and N.B. Zhang for helpful discussions. This work was supported in part by the U.S. Department of Energy, Office of Science, under Award No. DE-SC0013702.\\

\textit{DATA AVAILABILITY:} The data supporting the findings of this work,
including processed data and posterior samples, will be made openly
available \cite{Dataset}.


\vspace{1cm}
{\bf End Matter}--
\textit{The EOS model used and properties of cold matter Bayesian inferred from GSI proton flow data}

\begin{figure}[thb]
\centering
\vspace{-0.2cm}
 \resizebox{0.5\textwidth}{!}{
\includegraphics[width=1.\textwidth]{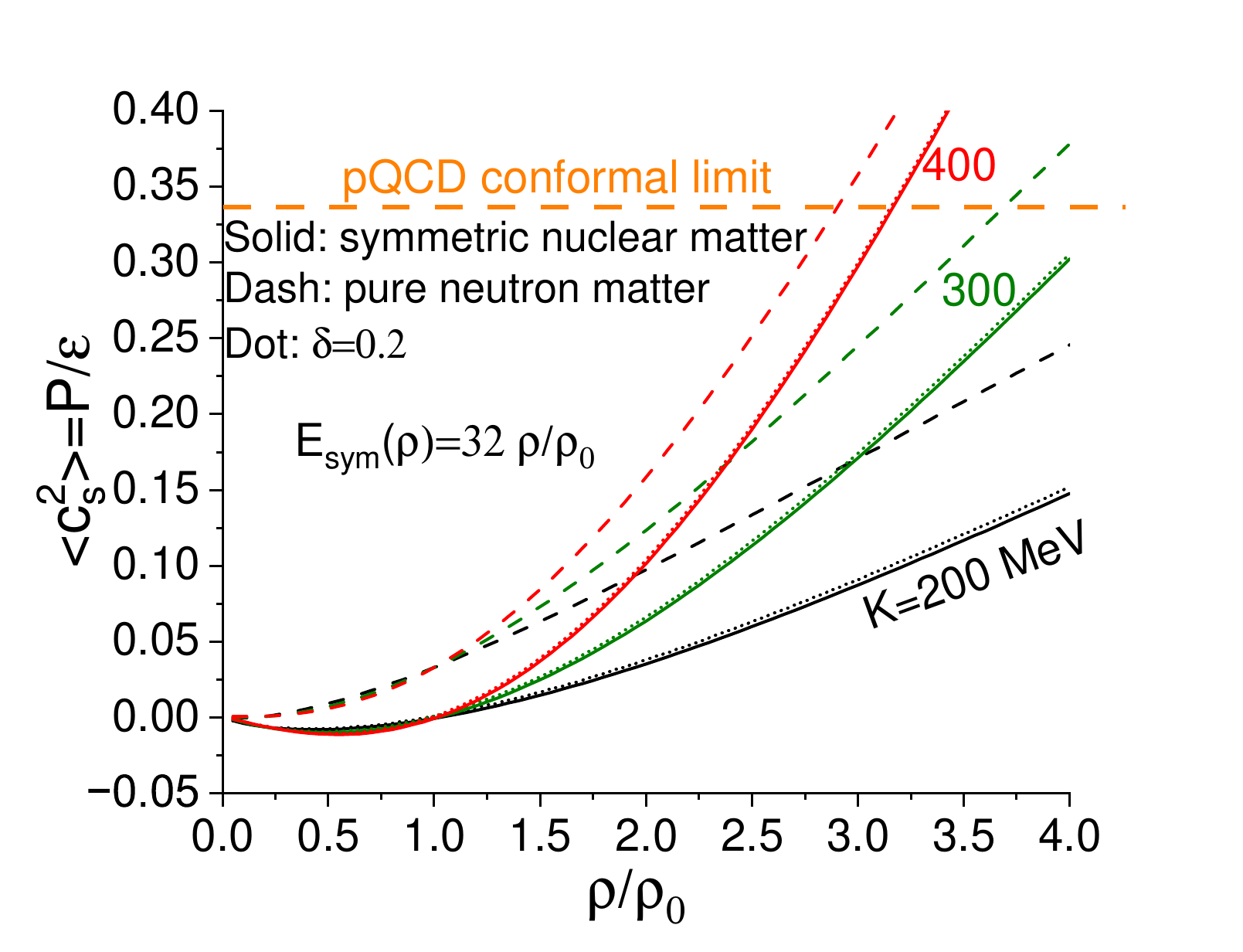}
}
 \resizebox{0.5\textwidth}{!}{
\includegraphics[width=1.\textwidth]{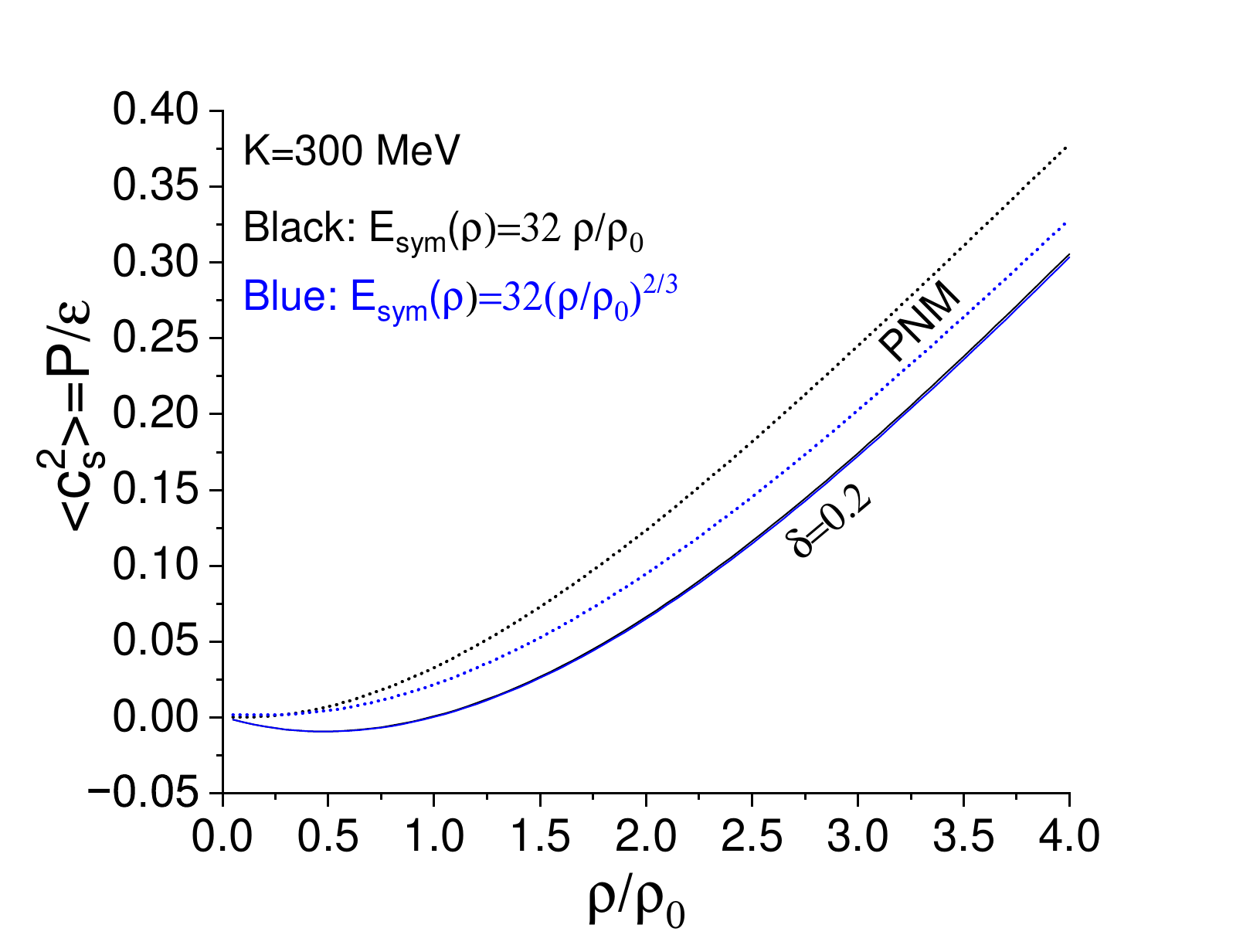}
}
\caption{The averaged speed of sound squared $<c^2_s(\rho)>=P/\varepsilon=w$ as a function of reduced baryon density with a Skyrme-like EOS with varying incompressibility parameter $K$ (upper panel) and nuclear symmetry energy (lower panel), respectively. 
} \label{phi12}
\end{figure}

The mean-field potential for a baryon $q$ is parameterized in a
Skyrme-like form
\begin{equation}
V_{q}(\rho,\delta) =
a \left(\frac{\rho}{\rho_0}\right)
+ b \left(\frac{\rho}{\rho_0}\right)^{\sigma}
+ V_{\rm sym}^{q}(\rho,\delta)
+ V^{q}_{\rm Cou},
\label{vq}
\end{equation}
where $V_{\rm sym}^{q}(\rho,\delta)$ denotes the symmetry potential in
isospin-asymmetric nuclear matter with isospin asymmetry
$\delta=(\rho_n-\rho_p)/\rho$, and $V^{q}_{\rm Cou}$ is the Coulomb
potential. The parameters $a$, $b$, and $\sigma$ are fixed by the saturation
properties of symmetric nuclear matter (SNM).
By varying the single incompressibility parameter
\begin{equation}
K = 9\rho_0^2
\left[\frac{d^2 E(\rho,\delta)}{d\rho^2}\right]_{\rho_0,\delta=0},
\end{equation}
one systematically scans the full stiffness range of the SNM EOS at
suprasaturation densities.

The corresponding energy per nucleon $E(\rho,\delta)$ in isospin-asymmetric nuclear matter is given by
\begin{equation}\label{EN}
E=M_N+\frac{3}{5}E_F^0\left(\frac{\rho}{\rho_0}\right)^{2/3}+\frac{a}{2}\frac{\rho}{\rho_0}
+ \frac{b}{1+\sigma}\left(\frac{\rho}{\rho_0}\right)^{\sigma}
+ E_{\rm sym}(\rho)\,\delta^2,
\end{equation}
where $E_F^0$ is the Fermi energy of a free nucleon gas at $\rho_0$ and
$E_{\rm sym}(\rho)$ is the nuclear symmetry energy.
In analyzing the GSI flow data, we adopt a conservative prior range
$K=180$--400~MeV.
As shown in the upper panel of Fig.~\ref{phi12} from SNM to pure neutron matter (PNM), for a typical isospin
asymmetry $\delta\simeq 0.2$ achievable in heavy-ion collisions and a
linear $E_{\rm sym}(\rho)$ motivated by relativistic mean-field models \cite{Chuck,Dut14},
the ratio $w=P/\varepsilon$ obtained with $P(\rho)=\rho^2\,dE(\rho,\delta)/d\rho$, spans a wide range at
suprasaturation densities.

Moreover, as illustrated in the lower panel of Fig.~\ref{phi12},
varying the symmetry energy from
$E_{\rm sym}(\rho)=32(\rho/\rho_0)^{2/3}$
(with slope parameter $L=64$ MeV, curvature
$K_{\rm sym}=-64$ MeV, and
$E_{\rm sym}(2\rho_0)=51$ MeV)
to $E_{\rm sym}(\rho)=32(\rho/\rho_0)$
($L=96$ MeV, $K_{\rm sym}=0$, and
$E_{\rm sym}(2\rho_0)=64$ MeV)
induces only a weak effect on $w$,
reflecting the dominance of the symmetric nuclear matter EOS in
both the pressure and energy density.
The effect would be even smaller if a symmetry energy softer than
$32(\rho/\rho_0)^{2/3}$ were adopted.
We note that these variations lie within current uncertainties
of the nuclear symmetry energy for $\rho\leq 3\rho_0$
\cite{Wolfgang,Bill,Li26a}.

Finally, it is worth emphasizing that the absence of explicit momentum
dependence in the employed single-nucleon potential allows a cleaner decoupling of
the cold-matter trace anomaly from thermal effects during the collision dynamics.
As first demonstrated in Ref.~\cite{Gale87}, if the momentum
dependence of the single-particle potential is included, it contributes
an additional term to the pressure of cold matter arising from Fermi
motion that depends on the strength and range of the
momentum-dependent interaction. These quantities are still not well
constrained, and their inclusion couples the nucleon mean field to the
finite-temperature kinetic pressure during the collision. As a result,
the thermal and cold contributions to the pressure become intertwined,
making it difficult to extract the zero-temperature EOS accurately from heavy-ion
collision observables.

In contrast, using momentum-independent mean fields allows the pressure
associated with the cold EOS to be separated more cleanly from thermal
effects. As discussed in the main text, the inverse Bayesian inference of the trace
anomaly of cold dense matter from collective flow is largely insensitive
to microscopic composition and interaction details (see, e.g., the HVH
relation for $P/\varepsilon$), and instead constrains the macroscopic EOS
$P(\varepsilon)$ regardless of how it is constructed. For this reason,
employing momentum-independent mean fields provides a practical and
transparent way to extract the cold dense-matter trace anomaly from heavy-ion
collisions and to facilitate direct comparisons with constraints
obtained from neutron star observations. In fact, most of the existing nuclear constraints on $P(\varepsilon)$ for cold dense matter were obtained consistently in such a manner from forward-modeling of heavy-ion collisions using various transport models \cite{Fuchs05,lynch09,Pawel02,LeF16,Cozma24,Jerome}.

\begin{table}
\centering
\caption{Beam energy dependence of the most probable $K$ and $X$ values inferred from the Bayesian analyses of FOPI's excitation functions of protons' directed and elliptic flow data \cite{Li25-PRC}.}
\begin{tabular}{cccc}\label{mean}
\\\hline
$E_{\rm{beam}}/A$ (MeV) &$K$ (MeV) & X\\\hline
150  &$220\pm 55$ & $0.7^{+1.1}_{-0.4}$\\
250 &$320\pm 62$ & $0.6^{+1.3}_{-0.5}$\\
400 &$313\pm 63$ & $0.9^{+1.3}_{-0.8}$ \\
600 &$312\pm 60$ & $0.9^{+1.4}_{-0.8}$\\
800 &$309\pm 63$ & $0.9^{+1.3}_{-0.7}$ \\
1000 &$310\pm 62$ & $0.9^{+1.3}_{-0.7}$\\
1200 &$323\pm 53$ & $1.0^{+1.4}_{-0.9}$\\
\hline
\end{tabular}
\vspace{-0.2cm}
\end{table}
Listed in Table~\ref{mean} are the most probable values of the incompressibility
$K$ and the in-medium scattering factor $X$ inferred from our Bayesian analyses
of the beam-energy dependence of protons' directed and elliptic flow \cite{Li25-PRC}.

The quoted $1\sigma$ uncertainties are obtained by marginalizing the posterior distributions of $K$ and $X$ and include experimental, emulator, and statistical uncertainties from the transport simulations \cite{LX-HA,Li25-PRC}. They do not include systematic effects associated with alternative transport implementations, EOS parameterizations, or prior choices, which may further broaden the error budget. Because broad, weakly informative priors are employed, the posterior distributions are primarily constrained by the data, making the inferred $\Delta(\varepsilon)$ largely insensitive to the specific prior choices.

\end{document}